\documentclass[aps,pre,twocolumn,float]{revtex4}
\usepackage{amsmath,bm,epsfig}
\newcommand{\beq}{\begin{equation}}
\newcommand{\eeq}[1]{\label{#1}\end{equation}}
\newcommand{\B}[1]{{\bm{#1}}}
\begin{document}
\title{Consequences of Disorder on the Stability of Amorphous Solids}
\author{Vladimir Dailidonis$^{1,2}$,  Valery Ilyin$^1$,  Pankaj Mishra$^1$ and Itamar Procaccia$^1$}
\affiliation{$^1$Weizmann Institute of Science,  Rehovot 76100, Israel \\
$^2$Bogolyubov Institute for Theoretical Physics, 03680 Kiev, Ukraine}
\begin{abstract}
Highly acurate numerical simulations are employed to highlight the subtle but important differences
in the mechanical stability of perfect crystalline solids versus amorphous solids. We stress the difference
between strain values at which the shear modulus vanishes and strain values at which a plastic instability insues. The temperature dependence of the yield strain is computed for the two types of solids, showing
different scaling laws: $\gamma_{_{Y}}\simeq \gamma_{_{Y}}^{0}-C_1 T^{1/3}$ for crystals versus $\gamma_{_{Y}}\simeq \gamma_{_{Y}}^{0}- C_2 T^{2/3}$ for amorphous solids.
\end{abstract}
\maketitle
\section{Introduction}
\label{intro}
It is well known that the mechanical stability of bulk crystalline solids at finite temperatures is dominated
by the motion of topological defects like dislocations. In perfectly ordered crystalline solids there are no
dislocations, and also in amorphous solids the notion of a dislocation does not exist since there is no long
range order with respect to which a dislocation can be defined. Both crystalline and amorphous solids resist
a small external stress (or strain) and return to their original shape
when the stress is removed. On the other hand, when higher stresses are applied some brittle solids break
while other ductile solids exhibit plasticity; they deform and do not return to their original shape when the stress
is removed.

Characterizing the mechanical strength of a given solid requires an understanding of the values of external stress or strain
at which the solid becomes mechanically unstable. We will refer to the values of stress where instabilities occur as ``critical streses".
For practical purposes one is interested in the so-called yield stress $\sigma_{_{Y}}$ which is defined as
the highest value of the stress which a solid can sustain before undergoing unbounded plastic flow. In a generic crystalline solid the yield stress  depends on the existence of defects, on temperature, on the time of the observation, etc.  Therefore, in order to define a sharp characteristic yield-stress one defines the ideal strength - the maximum achievable stress of a defect-free crystal at zero temperature.
The first attempt to estimate this value  for an ideal crystal which is elastically unstable was made by Frenkel \cite{F26}, Cf. Eqs. (\ref{FrenkU})-(\ref{Frenk}) below.
Recently it was shown \cite{07SES} that a crystal
can loose stability before the critical point predicted by Frenkel, i.e. when one vibrational mode reaches zero frequency. In fact, this loss of stability occurs {\it before} the shear modulus of the crystal vanishes. In this paper we will argue that one major consequence
of the randomness in amorphous solids is that the instability associated with the appearance of a soft vibrational mode (zero frequency)
is generically {\it after} the vanishing of the shear modulus. The reasons for this important difference will be elucidated and explained
in Sections \ref{hex} and \ref{glass}.

The critical stresses are calculated at zero temperature under quasistatic conditions as is explained
in Sect.~\ref{Models}.
In contrast, experiments are usually carried out at
finite temperatures.  Therefore it is important to extend the calculation of the critical stresses to finite temperatures.
In both perfect crystals and amorphous solids the values of the critical stresses reduce when the temperature is increased, simply
because it becomes easier to overcome the energy barrier involved in the mechanical instabilities. Nevertheless we will show below,
cf. Sect. \ref{glass}, that the difference between perfect crystals and amorphous solids translates to different temperature dependence
in the reduction of the critical stresses.

Sect. \ref{summary} presents a summary and conclusions of the present paper.


\section{Models and simulation methods}
\label{Models}

\subsection{Potentials}

In this section we introduce the numerical procedures that are common to our analysis of perfect crystals and amorphous solids.
The different implementations will be explained in subsequent sections.

In all our simulations we employ binary potentials between pairs of particles. In perfect crystals we have only one
type of particles, say $A$, and in the model amorphous solids we employ here two types of particles, say $A$ and $B$.
The interatomic interactions between particle $i$ (being A or B) and particle $j$ (being A or B) are defined  by shifted and smoothed Lennard-Jones
potentials
\begin{equation}
\phi_{ij}(r)=\left\{
\begin{array}{ll}
\phi_{ij}^{LJ}(r)+A_{ij}+B_{ij}r+C_{ij}r^2
&\textrm{if $r\le R^{cut}_{ij}$,}\\
0&\textrm{if $r> R^{cut}_{ij}$,}
\end{array}\right.
\label{KA1}
\end{equation}
where
\begin{equation}
\phi_{ij}^{LJ}(r)=4\epsilon_{ij}
\bigg[\bigg(\frac{\sigma_{ij}}{r}\bigg)^{12}
-\bigg(\frac{\sigma_{ij}}{r}\bigg)^{6}\bigg].
\label{LJ}
\end{equation}
The parameters are taken from  Ref. \cite{KA94}. All the potentials given by
Eq.~(\ref{KA1}) vanish with two zero derivatives at distances
$R^{cut}_{ij}=2.5\sigma_{ij}$. The parameters of the smoothing part
and details of the interparticle interactions can be found in Ref. \cite{DIMP14}.
It is convenient to introduce reduced units, with $\sigma_{AA}$ being the unit of length and
$\epsilon_{AA}$ the unit of energy.

\subsection{The preparation of the initial configuration}
\begin{figure}
\centering
\epsfig{width=.38\textwidth,file=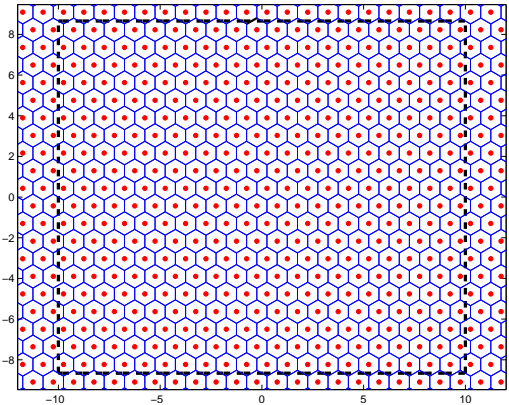}
\caption{Configuration of the one-component system with perfect hexagonal structure. The dotted lines represent the simulation box which is
continued periodically in both directions.}
\label{Fig1}
\end{figure}

The first step in all simulations is the construction of a model solid (crystalline or
amorphous) of $N$ particles in a two dimensional box of size $L_x\times L_y$ with periodic
boundary conditions. In the case of crystalline solid we place the $N$ particles on the vertices
of a hexagonal lattice, see for example Fig.~\ref{Fig1}.
Since the crystal is obviously free of defects it is also stress free. Thus the configuration is ready for subsequent straining.

The preparation of the amorphous solid is more involved. Firstly we equilibrate a system with 65\% particles A
and 35\% particles B at a temperature $T=1$ in Lennard-Jones units. This ratio is chosen to avoid crystallization
upon cooling. Next we cool the system to $T=10^{-6}$ in steps of $\Delta T=10^{-3}$ until $T=10^{-3}$ and then in one step
to the final temperature. The obtained configuration is not necessarily stress free, with particle
position denoted by $\B s_i$ from the set $\{\B s_i)\}_{i=1}^N$. Therefore we apply simple
shear which for a general strain $\gamma$ is defined by
\begin{equation}
\B r_i=\B h(\gamma)\cdot\B s_i \ ,
\label{ATJ1}
\end{equation}
with the transformation matrix
\begin{equation}
{\bf h(\gamma)}=\left(
\begin{array}{c c}
1& \gamma \\
0&1
\end{array}
\right).
\label{hG}
\end{equation}
Note that this transformation is volume preserving.

The configuration with (almost) zero stress is obtained at a strain $\gamma_0$;
the particle positions at this configuration are denoted by $\{\B r_i^0)\}_{i=1}^N$,
\begin{equation}
\B r^{0}_i=\B h(\gamma_0)\cdot\B s_i.
\label{ATJ2}
\end{equation}

Subsequently we strain the initial configuration, either crystalline or amorphous,  with additional
external affine simple shear. The procedure is as follows:
the particle positions change under shear strain from the reference state
$\{\B r^0_i\}$ to a
new one, denoted $\{\B r_i\}$, by an affine transformation that is defined
by a matrix ${\bf J}$ :
\begin{equation}
\B r_i=\B J\cdot\B r_i^0.
\label{ATJ}
\end{equation}
Here the matrix ${\bf J}$ in Eq.~(\ref{ATJ}) is given by
${\bf J}={\bf h}(\gamma)\cdot{\bf h}^{-1}(\gamma_0)$.
It follows from Eq.~(\ref{hG}) that the matrix $\B J$ is defined by
\begin{equation}
{\bf J(\gamma)}=\left(
\begin{array}{c c}
1& \gamma-\gamma_0 \\
0&1
\end{array}
\right),
\label{hG1}
\end{equation}
where the strain $\gamma_0$ corresponds to the deformation from the rectangular
simulation box to the reference system.

In the case of amorphous solid the affine transformation Eq.~(\ref{ATJ}) always destroys
the mechanical equilibrium. To regain mechanical equilibrium one should allow a non-affine
atomic-scale relaxation of the particle positions $\{\B r_i\}$ (see, e.g.,
\cite{LM06}). Also for a crystalline solid at finite temperature one should allow this step of non-affine
relaxation. At finite temperature this relaxation can be performed by
Molecular Dynamics or Monte Carlo methods. In the Monte Carlo protocol one moves the particles randomly and the move is
accepted with probability
\begin{equation}
P_{tr}=\min \bigg[1,\exp\bigg(-\frac{\Delta G}{T}\bigg)\bigg],
\label{trial}
\end{equation}
where $G$ is the generalized enthalpy. Under strain control the matrix
$\B h$ is fixed and the difference of the generalized enthalpy is defined by
the difference of the potential energy of the system
$U(\B h,\{\B s\})$
\begin{eqnarray}
\Delta G&=&U(\B h(\gamma),\B s_1,\cdots,\B s_i^{new},\cdots \B s_{N})- \nonumber \\
& &U(\B h(\gamma),\B s_1,\cdots,\B s_i^{old},\cdots \B s_{N}),\hspace{4 mm} 1\le i\le N,
\label{DifU}
\end{eqnarray}
where the displacement of the particle positions is
defined by
\begin{equation}
\B s^{new}_i=\B s^{old}_i+\delta \B s,\hspace{4 mm} 1\le i\le N
\label{Rmove}
\end{equation}
with the periodic boundary conditions taken into account.
In this equation the $\alpha$ component of the displacement vector of a
particle is given by
\begin{equation}
\delta s^{\alpha}=\Delta s_{max}(2\xi^\alpha-1),
\label{ParDisp}
\end{equation}
where $\Delta s_{max}$ is the maximum displacement and
$\xi^\alpha$ is an independent random number uniformly distributed between 0
and 1.

It follows from Eq.~(\ref{trial}) and Eq.~(\ref{DifU})  that in the limit
$T\to 0$ only the configurations with decreasing energy are accepted, i.e.,
the Monte Carlo process should converge to one configuration with minimal
energy. In practice the direct minimization of the energy of a system at zero
temperature after every small increase in strain (the athermal quasistatic (AQS)
strain control protocol \cite{ML04,KLLP10}) is more effective than the
stochastic Monte Carlo method.

\section{Hexagonal lattice}
\label{hex}

\subsection{Thermodynamic instability}
\begin{figure}[!h]
\centering
\epsfig{width=.38\textwidth,file=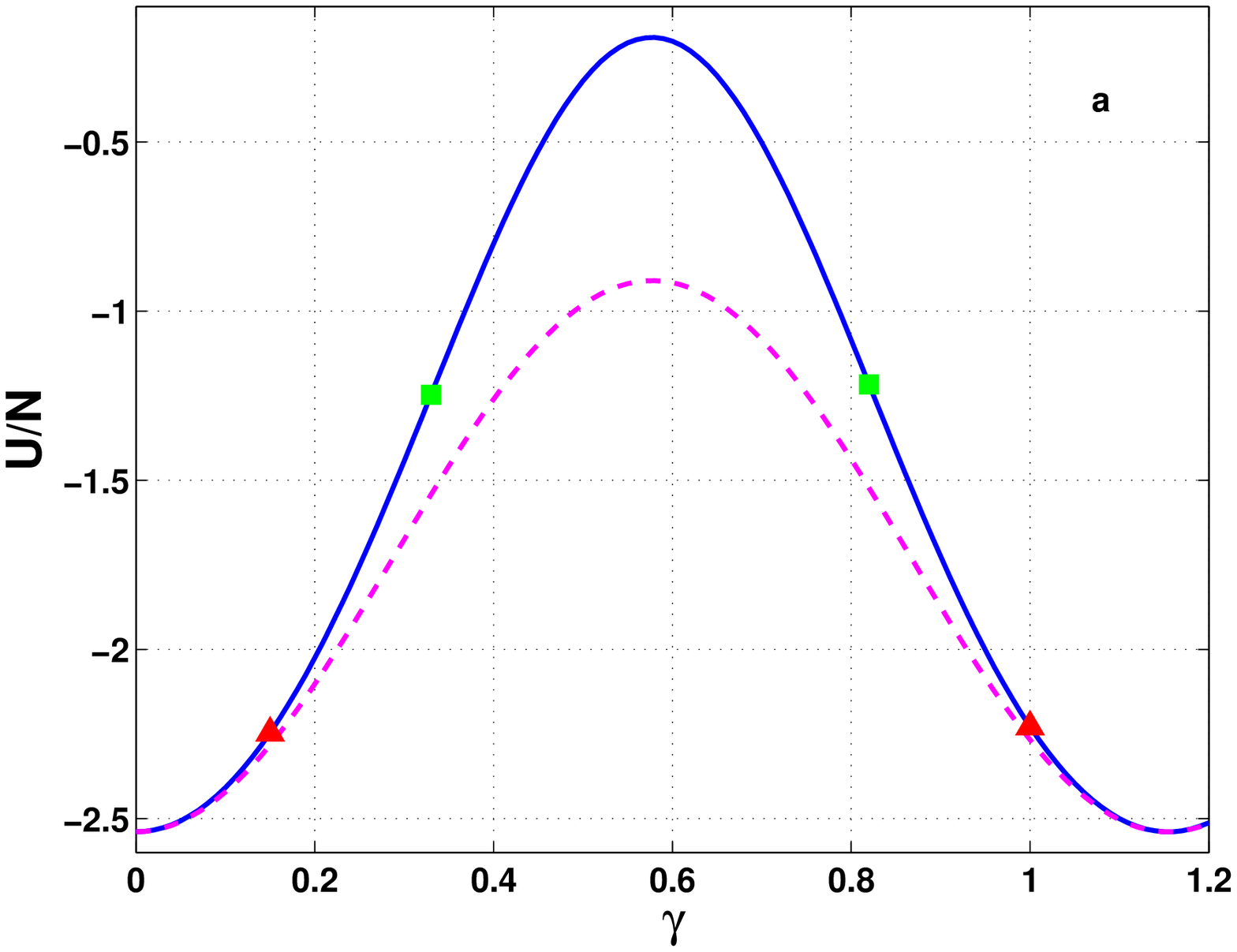}
\epsfig{width=.38\textwidth,file=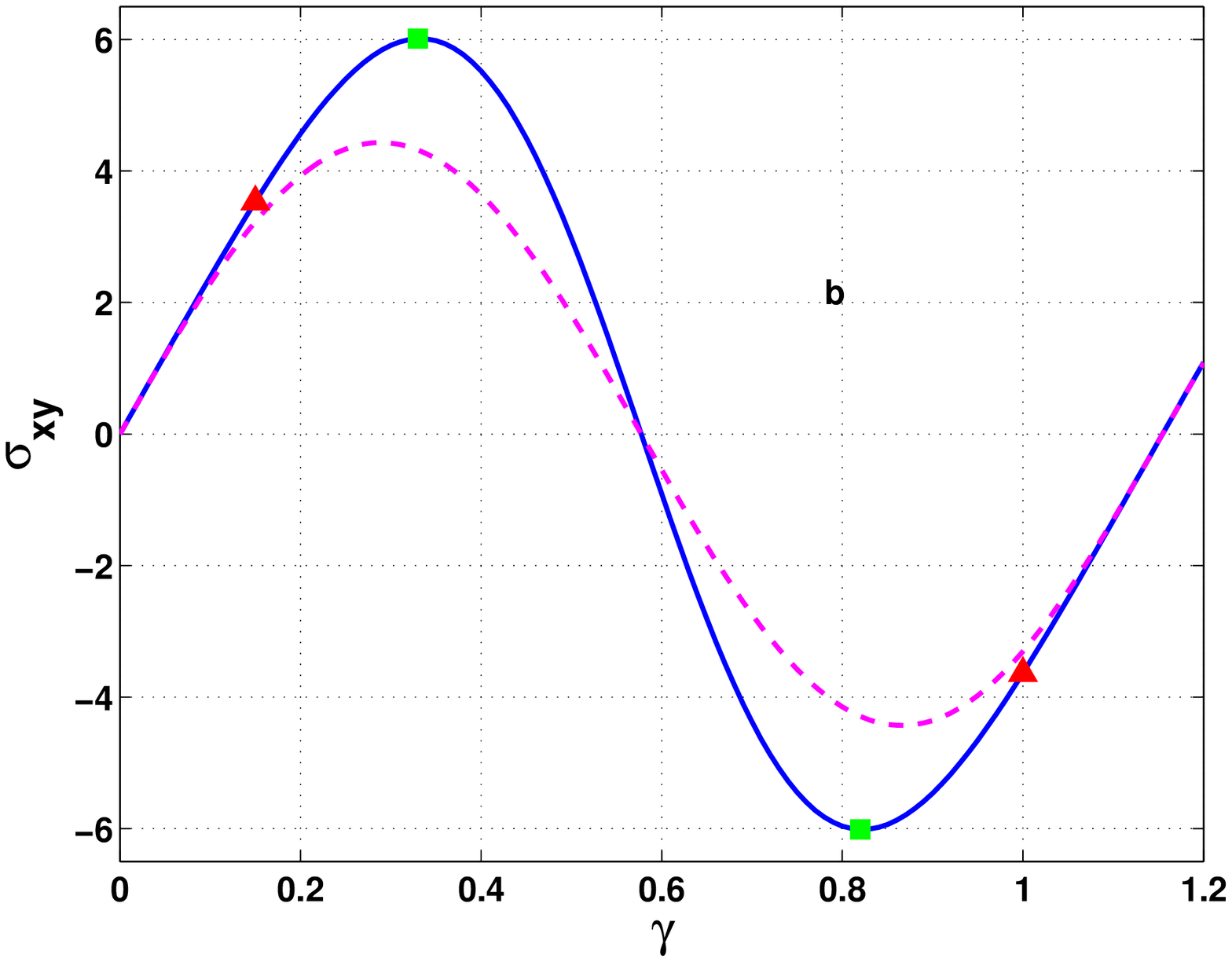}
\epsfig{width=.38\textwidth,file=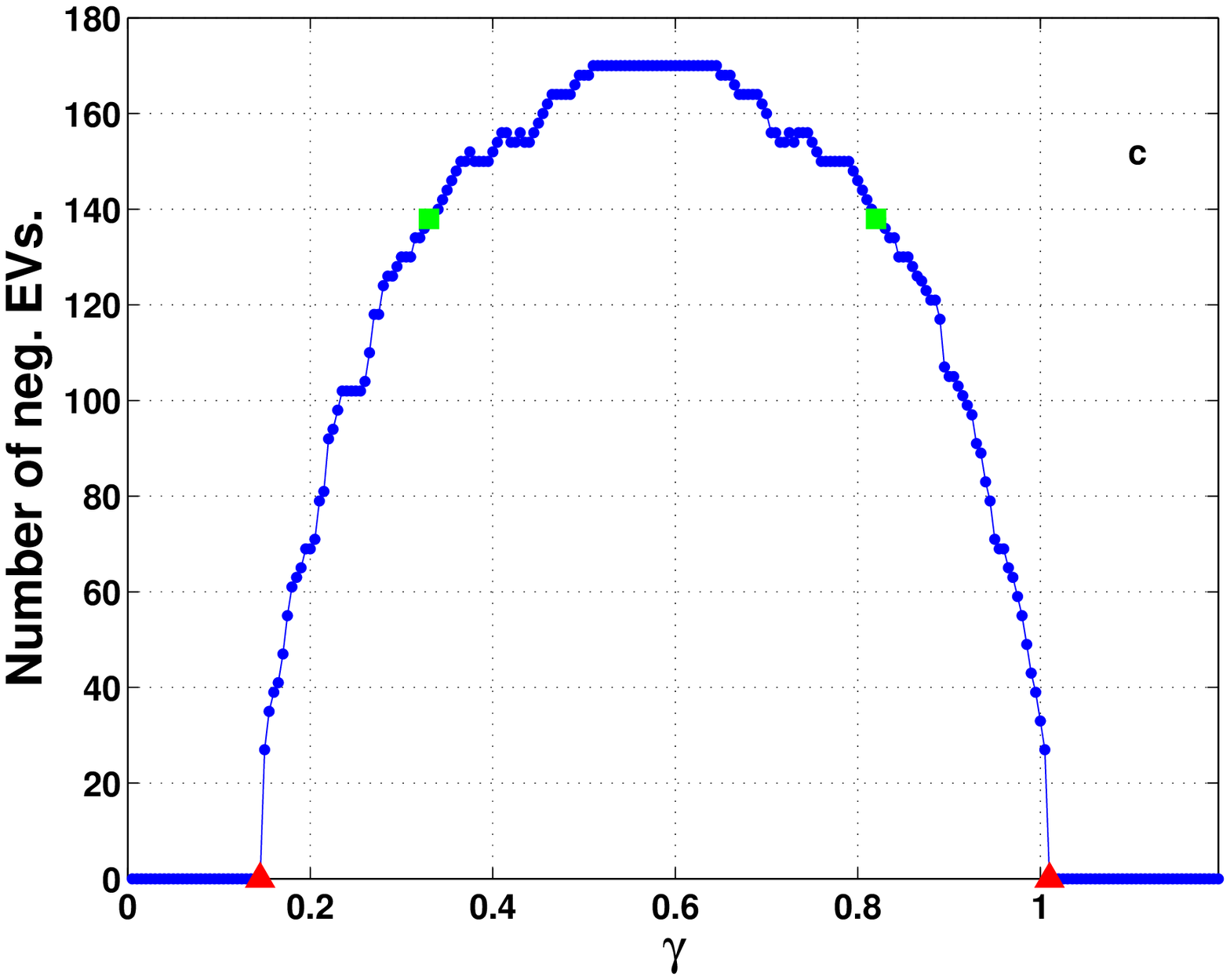}
\caption{The energy (a) and the shear stress (b) under simple shear. In blue continuous line we represent the exact, numerically
computed data. The dashed red line is the Frenkel approximation Eqs.~(\ref{FrenkU}) and (\ref{Frenk}). The red triangle and the
 green square represent the vibrational and the thermodynamic instabilities respectively. In panel (c)  we show the number of
negative eigenvalues of the Hessian for the system with $N=400$ as  a function
of the strain when nonaffine responses are suppressed by hand.}
\label{Fig2}
\end{figure}

The perfect hexagonal structure is shown in Fig.~\ref{Fig1}.
The energy of the system is minimal, $U/N=-2.5388472$, when the distance
between neighboring particles is $R_0=1.12152$ (at this point the pressure and
the internal shear stress are equal to zero) and the dimensionless particle number
density is $\rho=0.918$. The dependence of the
energy and the shear stress on the shear strain $\gamma$  under the simple shear defined by Eq.~(\ref{hG}) is shown
in Fig.~\ref{Fig2}. The elastic shear modulus of the system estimated at small
strains is equal $\mu=24.12$. Note that the shear modulus vanishes at the maximal and
minimal points of the stress vs. strain curve in the middle panel of Fig.~\ref{Fig2}.

\begin{figure}
\centering
\epsfig{width=.38\textwidth,file=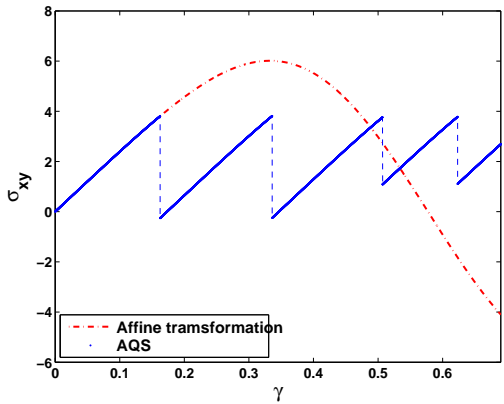}
\caption{Stress-strain relation for the perfect hexagonal lattice.
The solid line shows results of AQS simulatins, dotted line correspond to  the affine transformation (see also Fig.~\ref{Fig2}).}
\label{Fig2AA}
\end{figure}

The energy is a periodic function of the strain
and reaches its maximum when the hexagonal lattice is transformed into a square
one (which is unstable, see, e.g., \cite{FHP66}) at the strain $\gamma=1/\sqrt(3)$. It follows from the stress-strain
curve (middle panel) that the region between the points indicated by square symbols is
thermodynamically unstable. Frenkel proposed an analytical guess for the periodic functions
shown in Fig.~\ref{Fig2} in the form
\begin{equation}
U=\frac{\mu (1-cos(\sqrt{3}\pi\gamma))}{3 \pi^2}
 \label{FrenkU}
\end{equation}
and
\begin{equation}
 \sigma_{xy}^{int}=\frac{\mu}{\sqrt{3}\pi}sin(\sqrt{3}\pi\gamma) \ .
 \label{Frenk}
\end{equation}
The Frenkel approximation is shown in Fig.~\ref{Fig2} by the dashed lines. Both the approximation
and the numerical results indicate that the
stress can not exceed some value $\sigma^{int}_{xy}\le \sigma_{xy}^Y$. The quantitative
details differ.
Eq.~(\ref{Frenk}) yields the estimation
$\sigma_{xy}^Y=\mu/(\sqrt{3}\pi) \approx \mu/5$, underestimating
the result of direct numerical calculation
$\sigma_{xy}^Y\approx \mu/4$. In fact,  Eq.~(\ref{FrenkU}) and Eq.~(\ref{Frenk})
should be considered as first terms in a Fourier expansion \cite{L55}. The maximum value of the stress in the
approximation given by Eq.~(\ref{Frenk}) corresponds to the inflection
point of the strain-energy curve at $\gamma_Y=1/(2\sqrt{3})$ which is
associated with theoretical (ideal) strength which  is achieved by a
homogeneous deformation.

\subsection{Vibrational instability}

\subsubsection{Pure affine straining}

In fact, it is possible to lose stability during purely affine straining due to inhomogeneous
deformations by vibrational modes {\it before} becoming thermodynamically
unstable. The signifiers of such an instability are the eigenvalue of the Hessian matrix. At low temperatures the energy of a system in
the solid state can be written in the harmonic approximation
\begin{equation}
 U=U_0+ \Delta r_i^\alpha H_{ij}^{\alpha\beta}\Delta r_j^\beta\ ,
 \label{Harm1}
\end{equation}
with repeated indices summed upon and $\alpha$,$\beta$ denoting the cartesian components.
Here $U_0$ is the energy of a system in equilibrium
and the Hessian ${\B H}$ is the matrix
\begin{equation}
H^{\alpha\beta}_{ij}=\frac{\partial^2 U}{\partial r^\alpha_i \partial r^\beta_j}.
 \label{Hess}
 \end{equation}

In a canonical form Eq.~(\ref{Harm1}) reads
\begin{equation}
U=U_0+\sum\limits_{i}\lambda_i S_i^2,
\label{Harm2}
\end{equation}
where $\lambda_i$ are eigenvalues of the Hessian and $S_i$ are normal
coordinates. It follows from Eq.~(\ref{Harm2}) that in the harmonic approximation
a solid can be expressed as a number of uncoupled oscillators. The structure is stable
for arbitrary $S_i$ if all eigenvalues are positive. The unstable deformation
begins when the smallest eigenvalue approaches zero \cite{L98,ML99,MK00,GPL01,CKCM03,KUT04}.

The first eigenvalue $\lambda_P$ crosses zero {\it before} the shear modulus vanishes, at the value of
strain $\gamma_P$ denoted with the red triangle in Fig.~\ref{Fig2}. Note that when the strain increases this
eigenvalue becomes negative, and other eigenvalues cross zero and add up to a group of negative eigenvalues.
The dependence of the number of the negative eigenvalues on the strain under affine transformation is
shown in Fig.~\ref{Fig2} lower panel. The hexagonal lattice loses its stability as a harmonic system much before
the loss of thermodynamic stability. The reader should note that in practice one would never observe this increase
in the number of negative eigenvalues since the system will respond to the instability with non-affine responses
that are studied next. Here such non-affine effects were suppressed by hand.

For the perfect crystal without defects we expect the Hessian to be an analytic function
of $\gamma$ at least until the point of instability. In other words, we can write
\begin{equation}
<\B \Psi_p|\B H|\B \Psi_p>\equiv \lambda_P =A(\gamma_P-\gamma)+ B(\gamma_P-\gamma)^2+\cdots \ ,
\label{ana}
\end{equation}
where $\B \Psi_P$ is the eigenfunction of the Hessian associated with the eigenvalue $\lambda_P$ that
vanishes when $\gamma\to \gamma_P$. The consequences of this analyticity assumption are explored below.

\subsubsection{Relaxational effects}

The picture obtained with purely affine straining is incomplete.
For more precise and detailed information it is necessary to take into account relaxational effects in which
the system responds to the vanishing of an eigenvalue with non-affine motion.
To this aim we apply to the same crystalline hexagonal solid an athermal quasi-static protocol
in which after every increase $\Delta \gamma$ in the affine strain we follow up with gradient energy
minimization to regain mechanical equilibrium \cite{ML04b}.

\begin{figure}
\centering
\epsfig{width=.38\textwidth,file=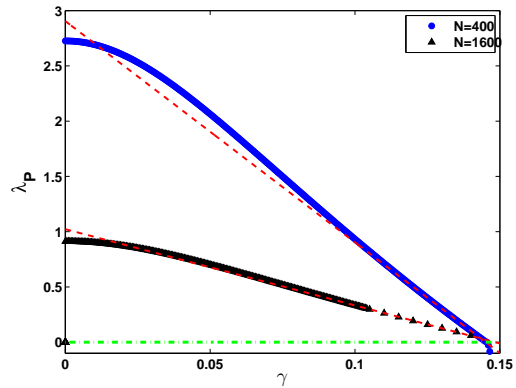}
\caption{Lowest eigenvalues of the Hessian for a perfect hexagonal lattice with particle number $N=400$ and $N=1600$
in the simulation box. The dashed red lines are an aid to the eye to observe the linearity of the dependence of the eigenvalue
on the strain.}
\label{Fig6}
\end{figure}

The strain-stress relation obtained in the frame of the AQS protocol is shown in Fig.~\ref{Fig2AA}.
One sees that the system loses stability before the point of the homogeneous instability. It is
useful to follow the trajectory of the lowest eigenvalue of the Hessian matrix as the strain is increased.
This is shown in Fig.~\ref{Fig6} for two system sizes with $N=400$ and $N=1600$.
The point at which the eigenvalue vanishes is the same for two system sizes. Near this instability
point the dependence of $\lambda$ on $\gamma$ is well represented by a linear law. This linearity is a direct
consequence of the analyticity assumption (\ref{ana}). This will be shown
to be in marked difference from the amorphous solid case.

\begin{figure}
\centering
\epsfig{width=.38\textwidth,file=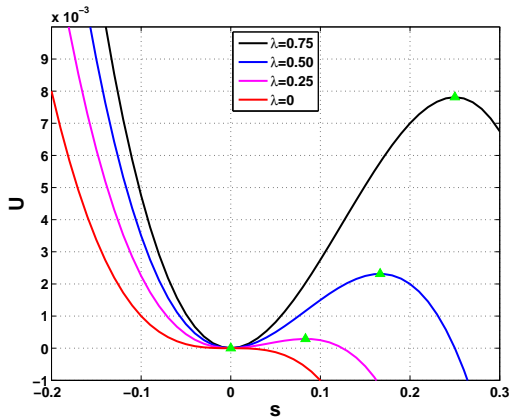}
\caption{Unharmonic model as given by Eq.~(\ref{anharm}). The green triangles denote
the extrema of the potential.}
\label{Fig5}
\end{figure}

When the harmonic approximation is being lost it is necessary to take
into account effects of anharmonicity in modelling the energy. The simplest model of an anharmonic
well is given by
\begin{equation}
U(s)=\frac{1}{2}\lambda_P(\gamma) S^2+\frac{1}{6}K S^3,
\label{anharm}
\end{equation}
where $\lambda_P(\gamma)$ is the lowest eigenvalue of the Hessian and $K$ is the constant
of the anharmonicity. The dependence of the energy given by Eq.~(\ref{anharm}) on the variable $S$
for different $\lambda_P(\gamma)$ is shown in Fig.~\ref{Fig5}.

It follows from Eq.~(\ref{anharm}) (see also Fig.~\ref{Fig5}) that the potential barrier is
related to the eigenvalue by
\begin{equation}
\Delta U(\gamma)=\frac{2}{3}\frac{\lambda_P(\gamma)^3}{K^2}
\label{barr}
\end{equation}

One should note that Eq.~(\ref{anharm}) is only approximate, taking into account only the most unstable mode.
In reality, especially in the thermodynamic limit, we expect other modes to intervene and dress the predictions
discussed above. This can be seen for example from the fact that the first instability shown in Fig.~\ref{Fig2AA}
occurs at $\gamma\approx 0.15$. On the other hand the eigenvalue $\lambda_P$ goes to zero at $\gamma\approx 0.14$.
Due to the intervention of other modes the eigenvalue should become ``slightly negative" before stability is actually lost.
To understand this further consider Eq.~(\ref{Harm2}). Upon the energy minimization after the affine step all eigenvalues
are effected, some of them increase and some decrease. The positive ones add to Eq. (\ref{Harm2}) positively and defer the
actual instability. If the energy minimization were performed precisely along the critical eigenfunction of the Hessian
this slight discrepancy would disappear.

\subsection{Monte Carlo studies at finite temperature}

Monte Carlo simulations are done at finite temperature, be it as small as it may. This blurs to some extent the
definition of the critical strains associated with the instabilities, since temperature fluctuation assist in
crossing potential barrier. Thus all the critical values discussed in this section should be understood as upper bounds.
It is always possible that longer Monte Carlo runs can result in lower value of the critical strains.

\begin{figure}
\centering
\epsfig{width=.38\textwidth,file=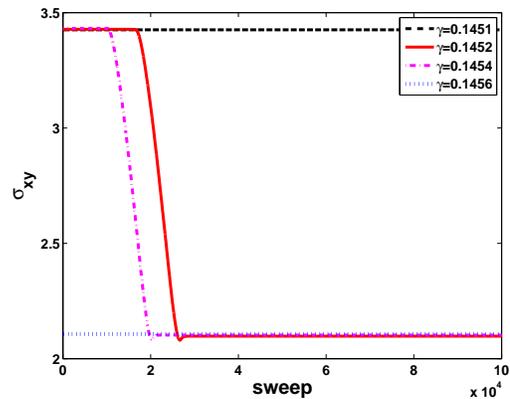}
\caption{Instantaneous values of the internal shear stress under strain control for different values of the applied strain.}
\label{Fig3}
\end{figure}

Instantaneous values of the internal shear stress under strain control Monte Carlo simulations
are shown in Fig.~\ref{Fig3}. For values of the strain less than some critical value the stress
fluctuates near a given average value. For some critical value of the strain the system
dwells for some time in a metastable state and then loses stability, transforming to a new stable state.
We chose the critical value of the strain corresponding to the appearance of metastable states.

\begin{figure}
\centering
\epsfig{width=.38\textwidth,file=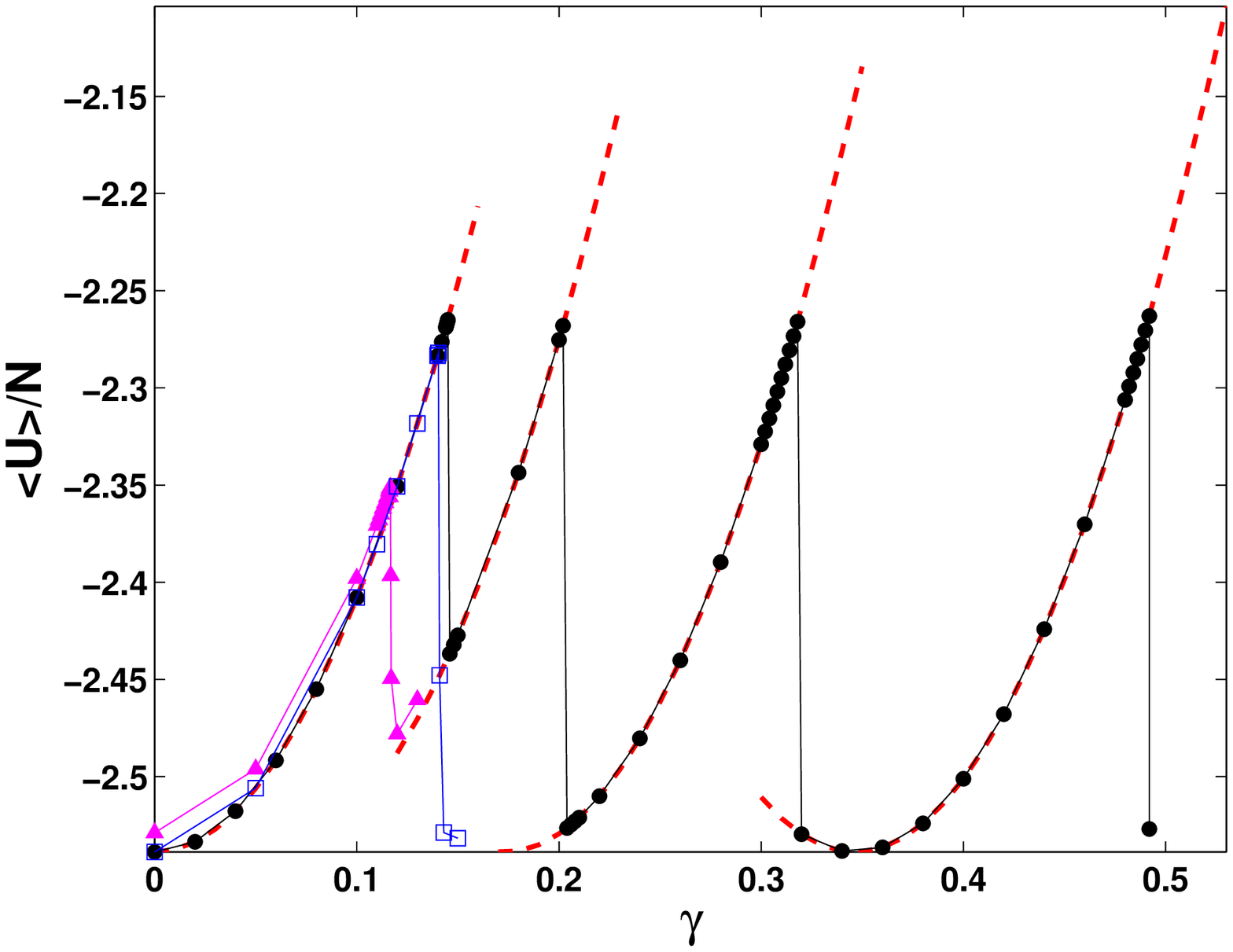}
\epsfig{width=.38\textwidth,file=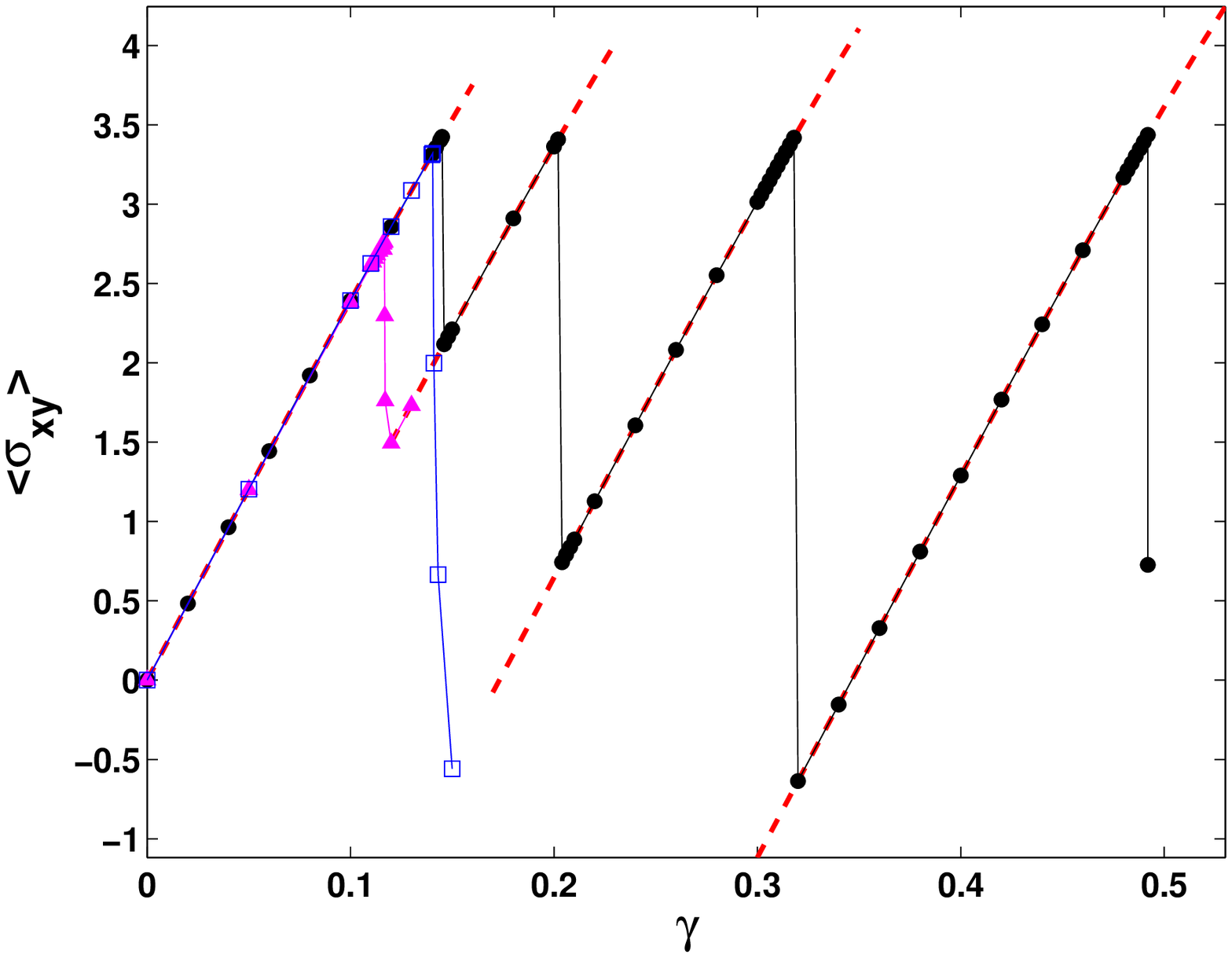}
\caption{The Monte Carlo results for the energy (upper panel) and the shear stress (lower panel) dependence on the strain for different temperatures.
Circles correspond to simulations at $T=10^{-6}$, squares to $T=10^{-4}$ and triangles to $T=10^{-2}$.}
\label{Fig4}
\end{figure}

Results of the Monte Carlo protocol for the mean values of the energy and
shear stress are shown for the crystal in Fig.~\ref{Fig4}. Under strain control the system undergoes a series of transitions associated with a loss of stability. Along each elastic branch the system follows the affine transformation (with the strain increased by some value $\gamma-\gamma_P$), see Fig.~\ref{Fig2}).
Each elastic branch is ending by a drop at different values of the strain but {\it with the same value of the energy and stress}. This values indicate the limit of the
stability of the hexagonal lattice. With increasing temperature the critical strains decrease.

At finite temperatures the barrier can be overcome if $T\sim \Delta U$, therefore, the
critical value of the eigenvalue is given by
\begin{equation}
\lambda_P(\gamma_P)\sim \bigg(\frac{3 K^2 T}{2}\bigg)^{1/3}.
\label{lambbarr}
\end{equation}
\begin{figure}
\centering
\epsfig{width=.38\textwidth,file=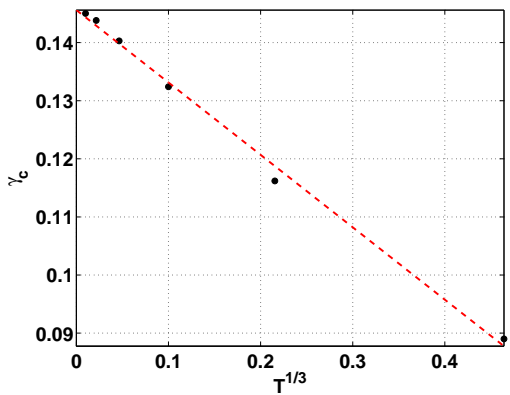}
\caption{Temperature dependence of the critical value of the strain for the perfect crystal.}
\label{Fig7}
\end{figure}
The dependence of the lowest eigenvalue of the Hessian (for two system sizes) on the strain estimated in the frame of
AQS is shown in Fig.~\ref{Fig6}.
The consequence of the analyticity assumption Eq.~(\ref{ana}) is that in the vicinity of the point $\gamma_P$ defined by $\lambda_P(\gamma_P)=0$ this dependence can be
approximated by the linear function $\lambda_P(\gamma)=A(\gamma_P-\gamma)$. Substitution of this expression to Eq.~(\ref{lambbarr})
yields
\begin{equation}
\gamma_{_{Y}}\simeq \gamma_{_{Y}}^{0}-C_1 T^{1/3}.
\label{DT13}
\end{equation}
Results of Monte Carlo indicate the correctness of this assessment (see Fig.~\ref{Fig7}).

\begin{figure}
\centering
\epsfig{width=.38\textwidth,file=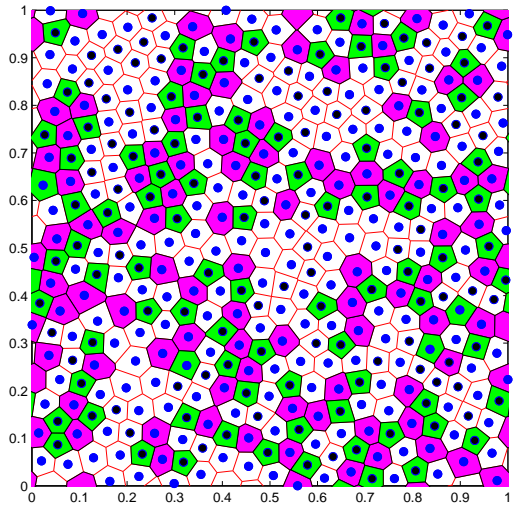}
\caption{Voronoi diagram for a glass configuration. The color code is green for pentagons, white
for hexagons and majenta for heptagons. Sometime an edge in the Voronoi cell can hard to visualize
at the scale of this image. }
\label{fig8}
\end{figure}

\section{Model glass}
\label{glass}
A composition of $A$ and $B$ particles that is stable in two-dimensions against
crystallization is chosen to be $65\%$ of particles $A$ and $35\%$ of particles $B$ \cite{BOPK09}.
The structure of the configuration of the binary mixture which produces our model glass is shown in Fig.~\ref{fig8}.

\begin{figure}
\centering
\epsfig{width=.38\textwidth,file=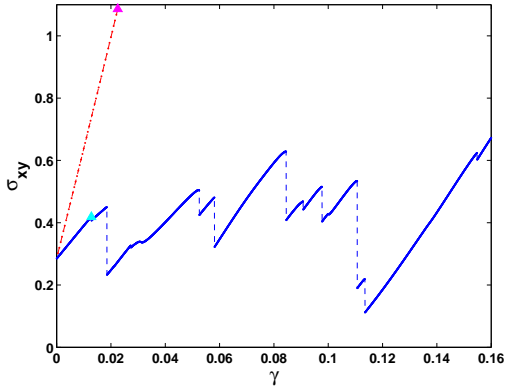}
\caption{AQS stress-strain relation for a glass. The serrated line corresponds to AQS simulations with non-affine corrections,
the dotted line shows stress-strain relation for a purely affine transformation (without non-affine corrections; the first points of instability is indicated by triangles.}
\label{fig9}
\end{figure}

The typical stress-strain relation of the model glass calculated in the frame of the AQS method is shown
in Fig.~\ref{fig9}. In contrast to the hexagonal lattice (see Fig.~\ref{Fig2AA}) instabilities are now appearing
at different values of the stress. This results from the fact that the hexagonal lattice
has only one reference state, in the glass there are many reference states and the transition between them
is caused by a saddle-node bifurcation that is accompanied by a sudden drop in stress.

\begin{figure}
\centering
\epsfig{width=.38\textwidth,file=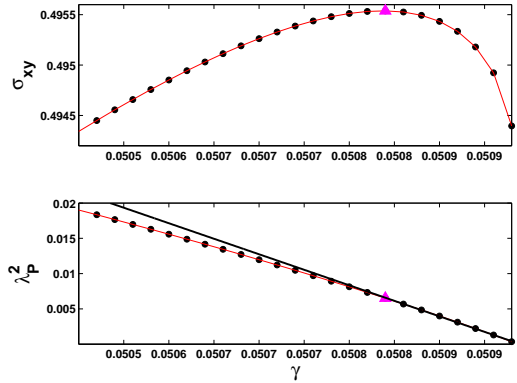}
\caption{The shear stress (upper panel) and lowest eigenvalue of the Hessian (bottom panel) dependence on the applied strain for
a glass configuration. Note that in this case the point A (denoted by the triangle) where the shear modulus vanishes {\em precedes}
point B where the Hessian lowest eigenvalue $\lambda_P$ goes to zero. }
\label{fig10}
\end{figure}

The fine structure of the stress-strain relation in the vicinity of the end of an elastic branch is shown in Fig.~\ref{fig10}.
One can see that there are two special points. One of them corresponds to the vanishing of the elastic modulus {\em followed} by the
instability point where the lowest eigenvalue of the Hessian goes to zero. It was shown in \cite{ML04} that the lowest eigenvalue of
the Hessian tends to zero as $\lambda_P\sim\sqrt{\gamma_P-\gamma}$, where $\gamma_P$ denotes the value of the strain at the instability point. When the system is not too large and the lowest eigenvalue
is well separated from the larger eigenvalues of the Hessian matrix it follows from this result (which is supported by the simulations) that
the elastic modulus in the critical region is approximated by
\begin{equation}
 \mu\approx \mu_B-\frac{A}{\sqrt{\gamma_P-\gamma}},
\label{MuC}
\end{equation}
where $\mu_B$ is the Born term. It follows from Eq.~(\ref{MuC}) that a theory for the glassy state in the spirit of the Frenkel approach would employ for the stress an analytic function in
the variable $x=\sqrt{\gamma_P-\gamma}$.  If applicable, the dependence of the stress on strain could be expanded in Taylor expansion around the point $\gamma_P$ \cite{KLLP10}
\begin{equation}
 \sigma_{xy}(\gamma)=\sigma_P+\sum\limits_{i=1}c_i (\gamma_P-\gamma)^{i/2},
 \label{TESS}
\end{equation}
where $c_2=\mu_B$. In fact this expansion may not exist and higher order term may diverge in the
thermodynamic limit due to the accumulation of small eigenvalues of the Hessian
(prevalence of many low lying barriers), as demonstrated in Ref.~\cite{11HKLP}

\subsection{The difference between crystal and glass}

Both for the hexagonal lattice and the glass there is a point of instability defined by a vanishing shear
elastic modulus (point A). Another instability point (point B), related to vanishing the lowest eigenvalue of the Hessian
appears {\em before} point A in the stress-strain dependence of the hexagonal lattice but {\em after} point A in the case of
glass. This difference has the following consequence: in the case of the hexagonal lattice when the strain is lower than
point A the system is thermodynamically stable, and there will be no important difference between stress-controlled and strain-controlled
protocols. In both cases the stress can be equilibrated in the system such that in stress-controlled protocols the internal and
the external stress are equal. Accordingly one can expect a similar temperature dependencies for $\gamma_{_Y}(T)$
under stress {\em or} strain control.

In contrast, in a glass under stress-control protocols the vanishing of the shear modulus
 is defined by point A with the lowest eigenvalue of the Hessian being still finite. Therefore, imagine that we apply to the glass a stress-controlled protocol with the external stress
being smaller than the critical stress at point A. At this situation the systems is still experiencing a barrier that needs to be overcome
since $\lambda_P\ne 0$. At $T=0$ therefore we will not experience an instability.

\begin{figure}
\centering
\epsfig{width=.38\textwidth,file=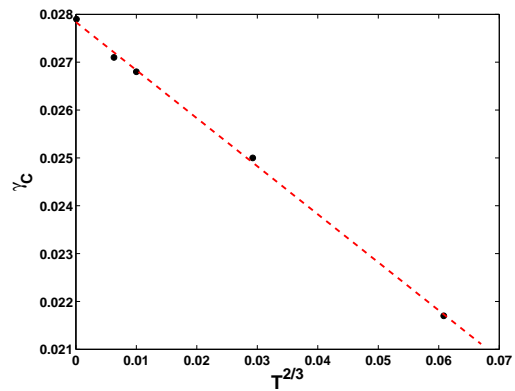}
\caption{Dependence of the critical strain value on the temperature for a glass.}
\label{fig11}
\end{figure}

The temperature dependence of the strain critical value obtained in the frame of the Monte Carlo protocol is shown in Fig.~\ref{fig11}.
The temperature  dependence of the yield strain is in agreement with $\sim T^{2/3}$ behavior \cite{JS05,DJHP13}.
\begin{figure}
\centering
\epsfig{width=.38\textwidth,file=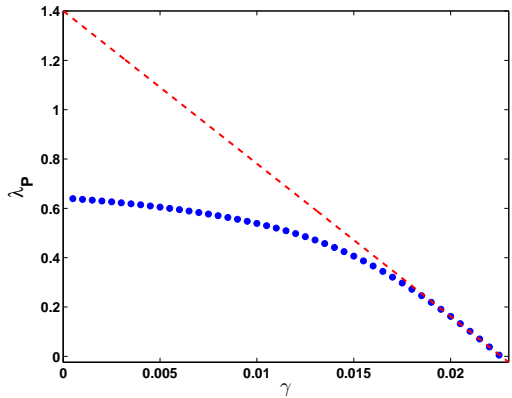}
\caption{The dependence of the lowest eigenvalue of the Hessian  on the applied strain for
a glass configuration under affine transformation.}
\label{fig13}
\end{figure}
\section{Conclusion}
\label{summary}

We have presented highly accurate numerical simulations to underline some fundamental
difference between the instabilities of glassy materials and perfect crystals, even when
the atomistic interaction are the same. The results indicate the importance of examining small
systems where the precise profiles of the stress vs. strain curves can be visualized. Increasing the
system size results in reducing the strain or stress differences between points of instability,
and eventually obliterating the details of the precise form of the stress vs strain characteristics.

Fundamentally, the difference is in the analytical dependence of the eigenvalues of the
Hessian matrix on the strain (or the stress). We note for example Fig.~\ref{fig9}, where
we highlight the distinction between straining the system allowing non affine response and not allowing it.
In the first case the eigenvalue has a square-root singularity as a function of the strain, as
discussed in Sect.~\ref{glass}. In the second case, cf. the dotted linear in Fig.~\ref{fig9}, the lowest eigenvalue of the Hessian matrix vanishes in an analytic fashion, liner in the strain, much in the same
way as in the crystalline case, cf. Fig~\ref{fig13}. The avoidance (by hand) of the saddle-node instability of the non affine response results in a fundamental change in the analytics of the dependence of the stress on the strain.


\end{document}